\documentclass[showpacs,prl,10pt,twocolumn,superscriptaddress,aps]{revtex4-1}

\usepackage{amsmath}
\usepackage{amssymb}
\usepackage{latexsym}
\usepackage{amsfonts}
\usepackage{epsfig}
\usepackage{psfrag}
\usepackage{graphicx}
\usepackage{wasysym}
\usepackage{ulem}
\usepackage{color}

\newcommand{\Eqref}[1]{Eq.~\eqref{#1}}

\begin{document}

\title{Magnetically amplified tunneling of the 3rd kind as a probe of minicharged particles} 

\author{Babette D\"obrich}
\altaffiliation{{Now at DESY, Notkestra\ss e 85, D-22607 Hamburg, Germany.}}
\author{Holger Gies}
\affiliation{Theoretisch-Physikalisches Institut,
Friedrich-Schiller-Universit\"at Jena, Max-Wien-Platz 1, D-07743 Jena, Germany}
\affiliation{Helmholtz-Institut Jena, Fr\"obelstieg 3, D-07743 Jena, Germany}
\author{Norman Neitz}
\altaffiliation{Now at Max-Planck-Institut f\"ur Kernphysik, Saupfercheckweg 1, D-69117 Heidelberg, Germany.}
\affiliation{Theoretisch-Physikalisches Institut,
Friedrich-Schiller-Universit\"at Jena, Max-Wien-Platz 1, D-07743 Jena, Germany}
\author{Felix Karbstein}
\affiliation{Theoretisch-Physikalisches Institut,
Friedrich-Schiller-Universit\"at Jena, Max-Wien-Platz 1, D-07743 Jena, Germany}
\affiliation{Helmholtz-Institut Jena, Fr\"obelstieg 3, D-07743 Jena, Germany}

\begin{abstract}
  We show that magnetic fields significantly enhance a new tunneling
  mechanism in quantum-field theories with photons coupling to
  fermionic minicharged particles. We propose a dedicated laboratory
  experiment of the light-shining-through-walls type that can explore
  a parameter regime comparable to and even beyond the best
  model-independent cosmological bounds. With present-day technology,
  such an experiment is particularly sensitive to minicharged
  particles with masses in and below the meV regime as suggested
    by new physics extensions of the Standard Model.
\end{abstract}

\date{\today}


\maketitle

Strong electromagnetic fields have recently become a powerful and topical
laboratory probe of fundamental physics \cite{Reviews}. Together with
precision optical probing, polarimetry experiments
\cite{Cameron:1993mr,Zavattini:2007ee} or experiments of the
light-shining-through-walls (LSW) type
\cite{Anselm:1986gz,Chou:2007zzc,Robilliard:2007bq,Fouche:2008jk,Afanasev:2008jt,Pugnat:2007nu,Battesti:2010dm,Ehret:2010mh} 
have provided 
the so far strongest laboratory -- and thus model-independent -- bounds on
axion-like particles (ALPs) or minicharged particles
(MCPs) \cite{Gies:2006ca}. Such hypothetical extremely weakly
  interacting particles occur
  in many new-physics models that are motivated by
  theoretical and observational puzzles in particle physics such as
  the strong-CP problem or dark-matter related anomalies. The
  particular power of laboratory experiments becomes obvious from the results of the 
  ALPS experiment  \cite{Ehret:2010mh}: In a parameter window near the meV
mass scale, ALPS provides for the most
stringent bounds on hidden-sector photons (further U(1) gauge bosons). The
maximum mass sensitivity scale of these experiments is typically set by the
frequency scale of the optical probe lasers $\sim$eV, such that these
experiments give access to a hypothetical new-physics regime of small masses
but very weak couplings, complementary to collider experiments. The underlying
mechanisms of induced polarimetric vacuum properties or photon-ALP conversion
yield observables which at best saturate for small masses as the mass parameter
effectively decouples in the small mass limit.

In the present work, we propose a search based on a new tunneling mechanism in
quantum field theory \cite{Gies:2009wx}: here a photon can traverse an
impenetrable barrier by virtue of virtual intermediate states that do not
couple to the barrier. As it complements standard quantum mechanical tunneling
and classical (tree-level) photon-ALP conversion, this phenomenon has been
dubbed ``tunneling of the 3rd kind''. It exploits the fluctuation-induced
nonlocal properties of quantum field theory. In principle, such a phenomenon
exists in the standard model with neutrinos as intermediate states, but the
effective photon-neutrino couplings are extremely weak due to the Fermi
constant \cite{Gies:2000tc}. For a search for new weakly
  interacting hypothetical particles,
this is a benefit as any standard-model-physics background is strongly
suppressed \cite{Ahlers:2008qc} compared with the signatures considered in
this work.

Whereas current laboratory bounds on minicharged particles are difficult to
improve with tunneling of the 3rd kind at zero field, we demonstrate here that an
external magnetic field can significantly amplify the tunneling probability
for the case of minicharged fermions. The essence of the
phenomenon lies in the existence of a near-zero mode in the Landau-type energy spectrum of
fermionic minicharged fluctuations. As this zero mode is screened only by the MCP
mass, the effect increases with a power-law dependence for decreasing MCP
mass or increasing magnetic field and approaches a maximum at the
pair-creation threshold.

Owing to this low-mass enhancement which is  unprecedented so far in the
context of strong-field physics, a dedicated laboratory experiment involving
only present-day technology has the potential to explore a parameter space
which so far had only been accessible with large-scale cosmological observations
based on CMB data \cite{Melchiorri:2007sq}, see also \cite{Ahlers:2009kh}. Astrophysical
considerations involving stellar energy loss arguments can even lead to stronger
MCP constraints \cite{Davidson:2000hf} but are somewhat model dependent
\cite{Masso:2006gc,Jaeckel:2006xm}. 

\begin{figure}
\includegraphics[width=0.35\textwidth]{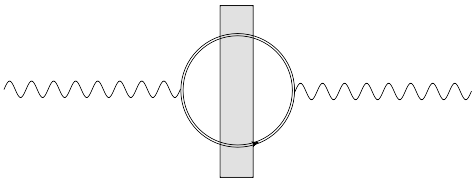} 
\caption{
Tunneling of a photon through a barrier mediated by a
  minicharged particle--antiparticle loop in a magnetic field.  While this
  process is also possible in a zero-field situation, cf. \cite{Gies:2009wx},
  it is considerably enhanced in a strong magnetic field indicated by the
  solid double line of the minicharged intermediate states.
}
\label{fig:t3k_full}
\end{figure}

The experimental tunneling setting resembles standard LSW setups, as sketched in
Fig.~\ref{fig:t3k_full}. A photon at frequency $\omega$ and momentum
$\mathbf{k}$ propagates orthogonally towards an opaque wall of
thickness $d$. The  system is put into a strong magnetic field
$\mathbf{B}=B\mathbf{\hat{e}_1}$ at an angle $\theta
=\varangle(\mathbf{B},\mathbf{k})$. A photon detector is placed behind
the wall.

We analyze the new tunneling phenomenon within an (effective) microscopic
quantum field theory with a QED-like Lagrangian, including a standard photon field $A_\mu$,
a Dirac spinor MCP $\psi_\epsilon$ (comments on scalar MCPs follow below),
and an interaction of the form
\begin{equation}
\label{lintDsp}
{\mathcal L}_{\rm int}= \epsilon\, e\,
\overline{\psi}_{\epsilon}
\gamma_\mu \psi_{\epsilon} A^\mu,
\end{equation}
where $\epsilon$ parametrizes the potentially small coupling strength in
units of the electron charge. The second unknown parameter of the theory is
the potentially light MCP mass $m$. As we expect the MCPs to remain
unobservable in direct measurements, we average over their fluctuations. This
leads to the effective Lagrangian for photon propagation in a strong
electromagnetic field
\begin{equation}
\mathcal{L}[A]= -\frac{1}{4} F_{\mu\nu} F^{\mu\nu}
- \frac{1}{2}\!\int_{x'}\!\!  A_\mu(x) \Pi^{\mu\nu}(x,x'|B) A_\nu(x'),\label{eq:calL}
\end{equation}
where $\Pi^{\mu\nu}(x,x'|B)$ denotes the photon polarization tensor in the
external field; here we have specialized to a magnetic field
$B$ which is assumed to be constant in all relevant spacetime regions,
implying translational invariance for $\Pi^{\mu\nu}$ to one-loop order.
Fluctuation-induced polarization effects of light propagating in a strong $B$
field that follow from this Lagrangian have been discussed in
\cite{Gies:2006ca,Ahlers:2006iz}. 
The associated equation of motion for the propagating photon in momentum
space reads ($k^2={\bf k}^2-\omega^2$)
\begin{equation}
 \left(k^2 g^{\mu \nu} - k^\mu k^\nu + \Pi^{\mu\nu}(k|B) \right) A_\nu (k) =0 \label{eq:EOM_Poltensor} .
\end{equation}
An important parameter in this context is provided by the strength of
the magnetic field relative to the MCP mass scale. The most relevant
regime for the present scenario is the strong-field domain, where
$\epsilon e B/ m^2\gg 1$. 
A particular enhancement of the tunneling effect occurs for the
Alfv\'{e}n-like transversal mode with polarization in
the $(\mathbf{k},\mathbf{B})$ plane. 
For non-vanishing $\theta$, this mode can be continuously related
to one of the transversal modes at zero field. The second (magneto-acoustic) polarization mode receives no dominant magnetic enhancement;
accordingly, the tunneling amplitude is not as strongly modified as for the Alfv\'{e}n-like mode.
Since photon propagation orthogonal to super-strong magnetic fields can be strongly 
damped \cite{Melrose:1976dr,dittrichgies,Shabad:1975ik}, we consider a
small angle between the direction of the magnetic field and the
propagation direction, $\theta\gtrapprox 0$. 
This is an important difference to standard LSW-type setups which typically employ $\theta=\pi/2$.
The equation of motion
for the Alfv\'{e}n-like transversal mode $A_{\text{T}}$ loses any nontrivial Lorentz
index structure, $\big(k^2 + \Pi(k) \big) A_{\mathrm{T}}(k)=0$, and
the polarization tensor for this mode to leading order in the $B$
field can be given as \cite{DobrichKarbstein,Karbstein:2011ja,Shabad:2003xy}
\begin{equation}
 \Pi(k)=\frac{\epsilon^2\alpha \epsilon
   eB}{2\pi}\,{e}^{-\frac{k_{\perp}^2}{2\epsilon eB}}\int\limits_{0}^{1}\!{\rm d}\nu
\frac{(1-\nu^2)\,k_{\parallel}^2}{m^2-i\eta+\frac{1-\nu^2}{4}
  k_{\parallel}^2},
\label{eq:Pi_fermion}
\end{equation}
where $k_\|=(\omega,k_1,0,0)$ and $k_\bot=(0,0,k_2,k_3)$ denote the
momentum components parallel and orthogonal to the $B$ field, and the
limit $\eta\to0^{+}$ is implicitly understood. 
Subleading corrections to \Eqref{eq:Pi_fermion} are at most logarithmic in $B$. Retaining the full
dependence on the photon momentum $k_\mu$ is essential here, since the
computation of the outgoing amplitude behind the wall requires to take
a Fourier transform back to position space. 
In the following, we assume reflecting boundary conditions at the wall for
the incoming photons, in agreement with the use of a cavity to enhance
the incoming photon flux. For the rear side, we assume absorbing
boundary conditions.
Other boundary conditions will lead to
slightly different prefactors $\sim \mathcal{O}(1)$ in our final
formulas. The probability for observing a photon at frequency
$\omega$ behind the wall via tunneling of the 3rd kind arises from
\begin{equation}
\label{eq:transition}
P_{\gamma\to\gamma}=\left|\int^{\infty}_{d} \!\!\!\!\!\mathrm{d} r'\, \frac{e^{-i\omega r'}}{2\omega}
\int^{0}_{-\infty} \!\!\!\!\!\mathrm{d} r''\, \Pi(r'\!-\!r'')\sin(\omega r'')\right|^2.
\end{equation}
{In principle} $r'{>d}$ runs over all points on the optical axis {between} the {rear side of the} wall {and the detector}, whereas $r''{<0}$ extends over all points {between the photon source and} the {front} side of the
wall.
Hence, the $r''$ integral samples all nonlocal contributions arising
from the incoming side and represents the source
for the outgoing photons. The $r'$ integral then coherently collects
all outgoing photons convoluted with the outgoing Green's function.
{However, in \Eqref{eq:transition} we have formally extended the respective integrations to $\pm\infty$. This is justified as $\Pi(r'\!-\!r'')$ receives its main contributions from relative distances $|r'\!-\!r''|$
of the order of the Compton wavelength $\sim1/m$ of the MCPs, and  falls off rapidly for $|r'\!-\!r''|\gg1/m$.
With respect to an actual experimental realization this implies that the magnetic field has to be sufficiently homogeneous only
within a sphere of diameter $\gtrsim1/m$ centered at the intersection of the optical axis with the wall.}  As we
consider the wall as perfectly opaque, we neglect here potential
couplings between the photons/MCPs and possible internal excitations of the
wall in a magnetic field.

The transition probability can in general be evaluated numerically
\cite{DGKN2}; analytic expressions follow from \Eqref{eq:Pi_fermion}
in various physically relevant limits. The full transition amplitude
is discussed in more detail in \cite{DGKN2}. In the following, we
concentrate on a specific set of conservatively chosen parameters
which can be experimentally realized with present-day technology.  For
the photon source and detection system, we consider state-of-the-art
parameters as successfully installed {and operated} at ALPS \cite{Ehret:2010mh}:
{The light of a frequency doubled standard laser light source, $\omega=2.33 \ \mathrm{eV}$ ($\lambda=532 \mathrm{nm}$),
is fed into an optical resonator cavity of length $L$ to increase the light power available for MCP production.
{So far, ALPS has employed $L\simeq 4$m, is currently upgrading to $L\simeq 10$m, but aims at $L\sim100$m in its 2nd state of expansion, ALPS-II.}  
Note that the divergence $\Delta\theta$ of the laser beam 
in an optical cavity is given by $\Delta\theta=\sqrt{\lambda/(\pi L)}$, i.e., for $L=10$m: $\Delta\theta=0.0075^{\circ}$
($L=100$m: $\Delta\theta=0.0024^{\circ}$).}

{The crucial difference to ALPS is the direction of the magnetic field, which in our scenario is at $\theta\gtrapprox 0$, instead of $\theta=\pi/2$.
As a suitable magnet we have identified a presently unused ZEUS compensation solenoid \cite{Dormicchi:1991ad} available at DESY.
It features a bore of $0.28$m diameter and $1.20$m length and provides a field strength of $B=5$T. The field points along the bore,
and is assumed to be adequately aligned on the solenoid's axis (accurate alignment studies of magnetic field lines relative to gravity have, e.g., been performed in \cite{Meinke:1990yk} for a HERA dipole magnet).
The field strength near the center of the solenoid is expected to be sufficiently homogeneous at least over a typical extent of the order of the bore's diameter.
The wall is installed in the center of the bore and the back end of the cavity extends into the bore. 
The angle $\theta$ is adjusted by tilting the entire optics assembly relative to the solenoid's axis.
Note that the detector position and its angular acceptance provides us with an additional handle to control $\theta$.}
On the one hand, a larger field strength enhances the discovery potential.
On the other hand, a sufficiently large spatial and temporal extent of the field is essential
for the sensitivity towards low-mass particles. {As discussed below \Eqref{eq:transition}}, the length
scale over which the field can be considered as approximately homogeneous
should be comparable to or larger than the Compton wavelength of the minicharged
particle. {Thus, with t}he ZEUS compensation solenoid, access to MCP masses
down to  $m \gtrsim 7\times 10^{-7}\mathrm{eV}$ is granted.

Even though our tunneling phenomenon -- contrary to LSW scenarios based on a tree-level process
-- intrinsically depends on the
thickness of the wall, this dependence turns out to be negligible in the small-mass/strong-field
limit which is of central interest here. We have checked that all our
results presented below are valid up to at least $d=1.8$cm as used in
\cite{Ehret:2010mh}.

In addition to the zero-field limit treated in \cite{Gies:2009wx}, also the
perturbative weak-field limit can be worked out analytically. However, even
the leading-order correction to the transition amplitude $\sim(\epsilon eB/m^2)^2$ turns out
to be quantitatively irrelevant in comparison to the zero-field effect for the
present parameters. Moreover, the accessible minicharged parameter space where
the perturbative expansion is valid is already ruled out by PVLAS data and
cosmological bounds. It is the nonperturbative strong-field limit of the
transition probability which gives access to a new region in the
particle-physics parameter space. Here, a characteristic scale is
provided by the condition for real pair creation
\begin{equation}
\omega \sin \theta \geq 2m. \label{eq:threshold}
\end{equation}
In the no-pair-creation (npc), strong-field regime
$\left\{\frac{\epsilon e B}{m^2},\frac{\epsilon e
  B}{\omega^2\sin^2\theta}\right\}\gg1$, the transition probability is
well approximated by 
\begin{equation}
P_{\gamma\to\gamma}^{\text{(strong,npc)}} \simeq \frac{\epsilon^4\alpha^2}{36 \pi^2}
\left( \frac{\epsilon e B}{m^2} \right)^2.
\label{eq:prob}
\end{equation}
This astonishingly simple asymptotics
can be understood in
a physical picture associated with the quantum fluctuations: the
typical length scale of the fluctuations is the Compton wavelength
$\sim 1/m$. It dominates all other length scales $\sim d$ and $\sim
1/\omega$ here, rendering the transition probability $d$ and $\omega$
independent in this regime. Equation \eqref{eq:prob} is also
independent of $\theta$ and thus represents the maximally available
transition probability in the limit $\theta\to 0$. 
However, for physically required finite values of $\theta$, real pair
creation eventually sets in if \Eqref{eq:threshold} is satisfied.

As real pairs are not expected to reconvert into photons, the
photon-tunneling probability will drop beyond the pair creation (pc)
threshold. In that strong-field regime with $\omega\sin\theta\gg2m$,
the transition probability for small angles $\theta$ becomes
\begin{equation}
 P^{({\rm strong,pc})}_{\gamma\to\gamma}\simeq \frac{\epsilon^4\alpha^2}{\pi^2} 
\,\frac{1}{\theta^8}
\left(\frac{\epsilon eB}{\omega^2}\frac{4m^2}{\omega^2}\right)^2 
\ln^2\left(\frac{2m}{\omega}\right),
\label{eq:strong_pc}
\end{equation}
and hence depletes with smaller mass but enhances with smaller
$\omega$. 

A prominent feature of Eqs.~\eqref{eq:prob} and
\eqref{eq:strong_pc} is the quadratic dependence on the magnetic
field, i.e., a linear dependence of the transition
amplitude on the parameter $\epsilon e B$.
This dependence leading to a small mass
enhancement in \Eqref{eq:prob} is a clear signature of IR dominance of the virtual
fluctuations. This IR dominance can be understood in terms of the
Landau level spectrum of virtual minicharged fluctuations in a
magnetic field. The eigenvalues of the squared Dirac operator for the
minicharged particles $\lambda_p = (p_\mu p^\mu)+m^2$ acquire the well-known
Landau-level structure in a magnetic field,
\begin{equation}
\lambda_{{p,j,\sigma}}=-p_0^2 + p_\|^2 + \epsilon e B (2j + 1 +\sigma) +
m^2.
\end{equation}
where $p_\|$ denotes the momentum component along the $B$ field, $j$ is the
Landau-level index and $\sigma=\pm 1$ labels the spin eigenvalues with respect
to the magnetic field. In the lowest Landau-level ($j=0$) and for $\sigma=-1$,
the eigenvalue reduces to a $1+1$ dimensional zero-field spectrum. This
dimensional reduction in quantum field theory goes along with an enhancement
of long-range fluctuations. The linear $B$-field dependence of the amplitude
then is dictated by the Landau-level measure in phase space. The long-range
fluctuations can finally be screened only by the Compton wavelength or
by real pair creation. 

Our proposed setup in an LSW experiment is the first that suggests to
exploit the characteristic near-zero mode of the spectrum  of
minicharged Dirac fermions. Other suggestions either work with photons near or
on the light cone \cite{Ahlers:2007qf}, with polarization properties
\cite{Gies:2006ca} or with fluctuation-insensitive thermal production rates as
in the case of cosmological bounds \cite{Melchiorri:2007sq}. These phenomena
are less sensitive to  minicharge masses and thus typically saturate in the
low-mass limit. 

An even stronger sensitivity arises near the pair creation threshold, where a resonance is encountered in the polarization tensor \cite{Shabad:1972rg}.
This resonance induces a
singularity in the transition amplitude in the idealized limit of
infinite coherent wave trains. If such resonances can
be exploited also for realistic finite wave packets, an even larger
parameter space could become accessible. In the present Letter we conservatively focus on the off-resonance regime, i.e., the parameter space that can be firmly excluded even if the encountered resonances would be smoothed out in an actual experimental realization.

The resulting observable in our setup is given by the outgoing photon rate on
the rear side of the light-blocking wall, 
\begin{equation}
\label{eq:conv}
n_{\mathrm{out}} = \mathcal{N} \  n_{\mathrm{in}} \  P_{\gamma\to\gamma} \ ,
\end{equation}
where $n_{\mathrm{in}}$ denotes the rate of incoming photons. The factor $\mathcal{N}$
accounts for a possible regeneration cavity on the rear
side of the wall. A feasibility study of this option even in the sub-quantum
regime was recently successfully performed in a dedicated experiment
\cite{Hartnett:2011zd}. In the absence of such a cavity, we have $\mathcal{N}=1$. 

\begin{figure}[h!]
\includegraphics[width=0.49\textwidth]{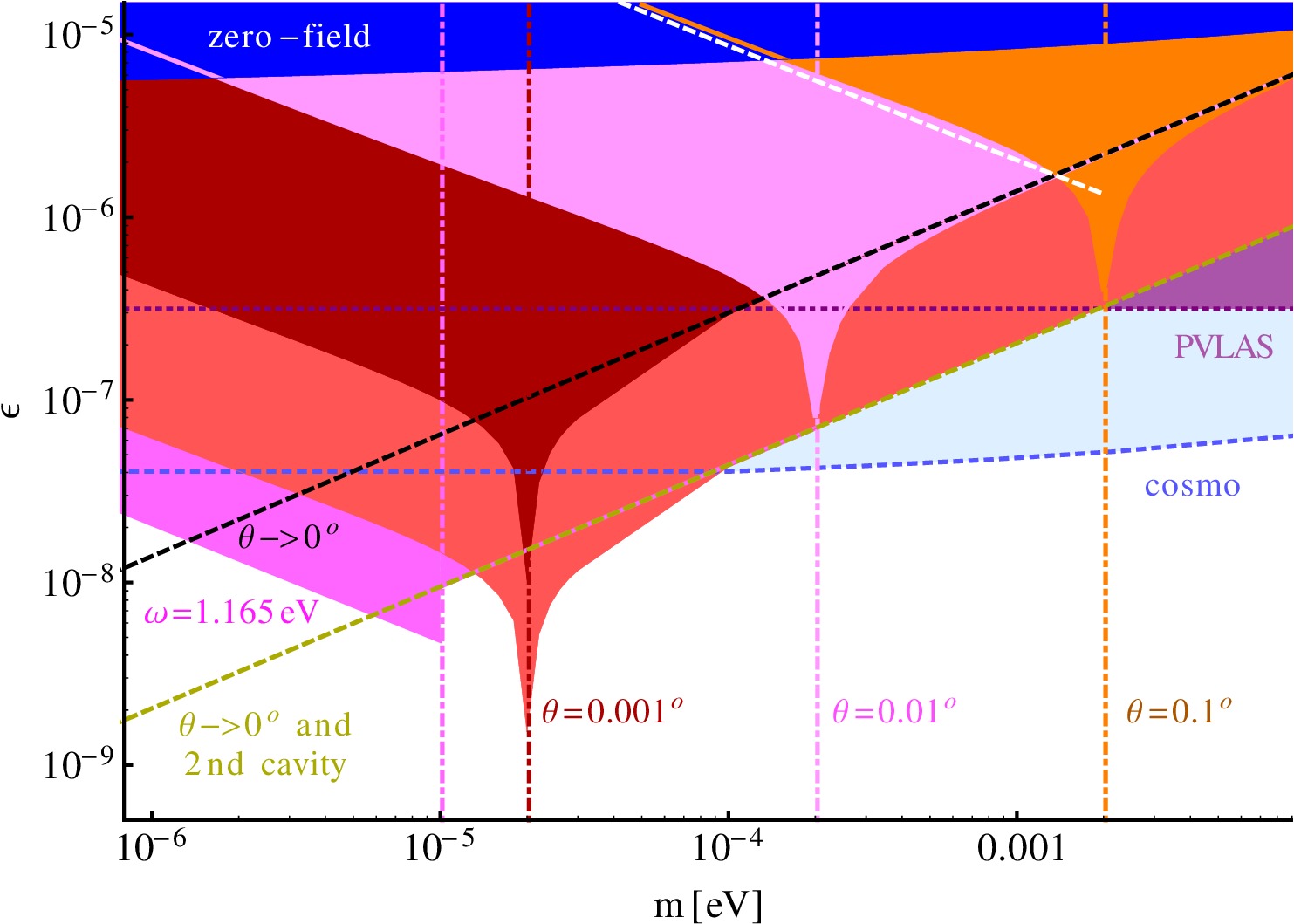} 
\caption{Accessible minicharge parameter space based on tunneling via
  virtual minicharged particles employing ALPS parameters \cite{Ehret:2010mh}. The transition at zero field gives
  access to the dark blue/dark shaded area, cf. \cite{Gies:2009wx}. The tunneling phenomenon
  is strongly amplified by a magnetic field (reddish/gray shaded areas),
  and can be further enhanced by the use of a 2nd cavity on the
  rear side (lowermost light red/gray shaded area using
  $\mathcal{N}=10^5$). The peak structures mark the resonance
  threshold \eqref{eq:threshold} for various angles $\theta$.
  {As discussed in the main text, the steep cusps at the pair creation thresholds might be smoothed and less pronounced in an actual experimental realization.}
  Our analytical estimate of \Eqref{eq:prob} without
  (black) and with (yellow/gray) additional cavity are shown as dashed
  lines and agree with the numerically computed
  strong-field limit. The asymptotics  \eqref{eq:strong_pc} beyond the pair creation
  threshold is indicated for one particular case as white dashed line.
  A comparison is made with limits
  \cite{Ahlers:2007qf} derived from PVLAS polarization measurements
  \cite{Zavattini:2007ee} (purple/dotted line), and the best
  model-independent cosmological bounds \cite{Melchiorri:2007sq}
  (blue/short-dashed line).  Our setup has the potential to
  outmatch these bounds with a regeneration cavity below $m\lesssim 9 \times 10^{-5}
  \mathrm{eV}$ with/without a fundamental mode laser at $\omega=1.165$eV
  (asymptotics indicated by the lower-left magenta area).
  }
\label{fig:excl_lhs}
\end{figure}

As demonstrated at ALPS \cite{Ehret:2010mh}, present-day technology
can achieve an incoming to outgoing photon ratio of
$n_{\mathrm{in}}/n_{\mathrm{out}}=10^{25}$, taking experimental issues
such as the effective detector sensitivity, run time and the use of a
front-side cavity into account. For the additional cavity on the
regeneration side, a factor of $\mathcal{N}=10^5$ appears realistic.
{A demanding issue with respect to an experimental implementation
  of our setting is the precise control of the angle $\theta$, which
  preferably should be very small, $\theta\gtrapprox0$. In
  Fig.~\ref{fig:excl_lhs} we present results for $\theta\geq
  0.001^{\circ}$.  As discussed above, the uncertainty in the
  adjustment of $\theta$ is expected to be dominated by the beam
  divergence $\Delta\theta$.  Notably, even with a divergence of
  $\Delta\theta=0.0024^{\circ}$ exclusion bounds of the same quality
  as presented in Fig.~\ref{fig:excl_lhs} for $\theta=0.001^{\circ}$
  should become experimentally viable: Due to the fact that $\theta$
  and $\Delta\theta>\theta$ are of comparable size, effectively both
  smaller and somewhat larger angles as $\theta=0.001^{\circ}$ are
  sampled.  Predictions for a concrete experimental set-up
    require, of course, a detailed modeling also including the profile
    of the cavity mode. However, we expect our present estimates to be
    affected only by prefactors of $\mathcal{O}(1)$. }

In Fig.~\ref{fig:excl_lhs}, we compare our resulting parameter space with current
experimental exclusion limits \cite{Ahlers:2007qf} based on PVLAS polarization
measurements \cite{Zavattini:2007ee} (purple/dotted line), and the best
model-independent cosmological bounds \cite{Melchiorri:2007sq} obtained from
CMB data (blue/short-dashed line).

To summarize: even with the conservatively chosen parameters, we find
that our magnetically amplified tunneling scenario can significantly
enhance the discovery potential for minicharged particles in an LSW
experiment. This setup can improve PVLAS polarization data for
minicharged particles below $m\lesssim 2 \times 10^{-4} \mathrm{eV}$.  Employing a cavity on the regeneration
side with/without a laser in the fundamental mode with
$\lambda=1064\mathrm{nm}$ ($\omega=1.165$eV), these values can even be
improved beyond PVLAS bounds for $m\lesssim 2 \times 10^{-3}
\mathrm{eV}$ and cosmological bounds below $m \lesssim 9 \times
10^{-5} \mathrm{eV}$.

As this mechanism of magnetic amplification is only active for Dirac fermionic
fluctuations due to the underlying Landau-level structure, our proposal can
decisively distinguish between minicharged scalars or fermions.

Finally, it
appears worthwhile to consider similar ideas also on
terrestrial or astrophysical scales, as magnetic fields of larger extent might give access to even further regions of the minicharged particle parameter space.

BD, HG and NN acknowledge support by the DFG under grants SFB-TR18 and
GI~328/5-2 (Heisenberg program) as well as GRK-1523.  We thank
J.~Jaeckel, A.~Lindner and D.~Trines for interesting discussions and
helpful correspondence.

\end{document}